\documentclass[final,journal]{IEEEtran}

\usepackage{cite}
\usepackage{amsmath}
\usepackage{graphicx}
\usepackage{stfloats} %to have equation across the bottom of the page

\usepackage{color}

\usepackage{balance}

\usepackage{tikz}

\usepackage{pgfplots}
\pgfplotsset{width=5cm,compat=1.9}
\usepgfplotslibrary{external}
\usepgfplotslibrary{statistics}

\usepackage{cite}
\makeatletter
\def\bstctlcite{\@ifnextchar[{\@bstctlcite}{\@bstctlcite[@auxout]}}
\def\@bstctlcite[#1]#2{\@bsphack
 \@for\@citeb:=#2\do{%
   \edef\@citeb{\expandafter\@firstofone\@citeb}%
   \if@filesw\immediate\write\csname #1\endcsname{\string\citation{\@citeb}}\fi}%
 \@esphack}
\makeatother

\begin{document}

\bstctlcite{IEEEexample:BSTcontrol}

\title{Probabilistic Scheduling of UFLS to Secure Credible Contingencies in Low Inertia Systems}
%Probabilistic Scheduling of UFLS to Secure the Largest Credible Contingency
%Probabilistic Scheduling of UFLS to Secure Credible Contingencies in Low Inertia Systems
% Other option: "Probabilistic Scheduling of Under-Frequency Load Shedding in Low-Inertia Systems"
% "Probabilistic" to explain that we are not considering UFLS as a market service where providers get paid every hour to be ready to disconnect (like in France or Italy in the European split of Jan 2021, they used industries that they have a commercial agreement with)
% "Largest Credible Contingency" means N-1

\author{Cormac~O'Malley,~\IEEEmembership{Student Member,~IEEE,}%
\ Luis Badesa,~\IEEEmembership{Member,~IEEE,}%
\ Fei Teng,~\IEEEmembership{Member,~IEEE,} \\ %
and Goran Strbac,~\IEEEmembership{Member,~IEEE}%
%\thanks{Cormac is an Imperial PhD Student}% <-this % stops a space
\vspace{-0.8cm}
\thanks{The authors are with the Department of Electrical and Electronic Engineering, Imperial College London, SW7 2AZ London, U.K. (email: c.omalley19@imperial.ac.uk).}
}

\markboth{IEEE Transactions on Power Systems, October 2021}%
{Paper Header 2}

\maketitle

\begin{abstract}
The reduced inertia levels in low-carbon power grids necessitate explicit constraints to limit frequency's nadir and rate of change during scheduling. This can result in significant curtailment of renewable energy due to the minimum generation of thermal plants that are needed to provide frequency response (FR) and inertia. Additional consideration of fast FR, a dynamically reduced largest loss and under frequency load shedding (UFLS) allows frequency security to be achieved more cost effectively. This paper derives a novel constraint from the swing equation to contain the frequency nadir using all of these services. The expected cost of UFLS is found probabilistically to facilitate its comparison to the other frequency services. We demonstrate that this constraint can be accurately and conservatively approximated for moderate UFLS levels with a second order cone (SOC), resulting in highly tractable convex problems. Case studies performed on a Great Britain 2030 system demonstrate that UFLS as an option to contain single plant outages can reduce annual operational costs by up to £559m, 52\% of frequency security costs. The sensitivity of this value to wind penetration, abundance of alternative frequency services, UFLS amount and cost is explored.
\end{abstract}


\section*{Nomenclature}
\subsection*{Indices and Sets}
\begin{itemize}[\settowidth{\labelsep}{HH} \settowidth{\labelwidth}{HELL}]
\item[$g,G$] Index, Set of generators.
\item[$k,K$] Index, Set of UFLS levels.
\item[$n,N$] Index, Set of nodes in the scenario tree.
\item[$s,S$] Index, Set of storage units.

\end{itemize}

\subsection*{Constants}
\begin{itemize}[\settowidth{\labelsep}{HH} \settowidth{\labelwidth}{HELL}]
\item[$\Delta f_{trig}$] Low frequency deviation at which UFLS is triggered (Hz).
\item[$\Delta \tau(n)$] Time-step corresponding to node $n$ (h).
\item[$\pi (n)$] Probability of reaching node $n$. 
\item[$c_{LS}$] Value of load-shed from lack of reserve (£/MWh).
\item [$c^{UFLS}_k$] Value of load-shed at UFLS level $k$ (£/MWh).
\item[$f_0$] Nominal grid frequency (Hz).
\item [$M$] Big-M constant.
\item [$N_{PL_{max}}$] Number of units generating at $PL_{max}$.
\item [$p$] Probability of $PL_{max}$ outage (occurrences/h).
\item[$P^d(n)$] Total demand at node $n$ (GW).
\item[$P^w(n)$] Total wind power availability at node $n$.
\item [$P^{UFLS}_{k}$] Load-shed amount at level $k$ (GW).
\item[$RoCoF_{max}$\hspace{-0.5cm}] \hspace{0.5cm} Maximum admissible RoCoF (Hz/s).
\item [$t_{rec}$] Disconnection length of UFLS (h).
\item [$T_i$] Delivery speed of FR type $i$ (s). 
\item [$T_s$] RoCoF sampling period (s).
\end{itemize}

\subsection*{Decision Variables (continuous unless stated)}
\begin{itemize}[\settowidth{\labelsep}{HH} \settowidth{\labelwidth}{HELL}]
\item[$H$] System inertia after the loss of $PL_{max}$ (GWs).
\item [$m_k$] Binary variable corresponding to UFLS level $k$.
\item[$P_g(n)$] Power output of units $g$ at node $n$ (GW).
\item[$P^{LS}(n)$] Load-shed from lack of reserve at node $n$ (GW).
\item[$P_s(n)$] Power output from storage $s$ at node $n$ (GW), positive discharge, negative charge.
\item[$P^{wc}(n)$] Wind curtailment at node $n$ (GW).
\item[$PL_{max}$] Largest power infeed (GW).
\item[$R_i$] Magnitude of type $i$ FR (GW).

\end{itemize}

\subsection*{Linear Expressions of Decision Variables}
\begin{itemize}[\settowidth{\labelsep}{HH} \settowidth{\labelwidth}{HELL}]
\item[$C_g(n)$] Operating cost of units $g$ at node $n$ (£).
\item[$C^{U}(n)$] Expected cost of UFLS at node $n$ (£).
\item[$R_i(t)$] Time evolution of FR type $i$ (GW).
\item[$X,Y,Z$] Auxiliary expressions for eq.~(\ref{non convex SOCP}).
\end{itemize}

\subsection*{Nonlinear Expressions of Decision Variables}
\begin{itemize}[\settowidth{\labelsep}{HH} \settowidth{\labelwidth}{HELL}]
\item[$\Delta f(t)$] Frequency deviation at time $t$ after outage (Hz).
\item[$\hat{t}$] Time after outage that frequency reaches the UFLS trigger level (s).
\end{itemize}

\subsection*{Acronyms}
\begin{itemize}[\settowidth{\labelsep}{HH} \settowidth{\labelwidth}{HELL.}]
\item[CCGT] Combined Cycle Gas Turbine
\item[FS] Frequency Services
\item[FSC] Frequency Security Cost
\item[FR] Frequency Response
\item[MISOCP] Mixed-Integer Second Order Cone Programme
\item[OCGT] Open Cycle Gas Turbine
\item[RoCoF] Rate of Change of Frequency
\item[SOC] Second Order Cone
\item[SUC] Stochastic Unit Commitment
\item[UFLS] Under Frequency Load Shedding
\end{itemize}

\section{Introduction}
\IEEEPARstart{G}{rid} frequency needs to be kept within boundaries to prevent system damage and emergency demand disconnection. The rate of change of frequency (RoCoF) is determined by the instantaneous generation-demand imbalance.
Post generator outage, the frequency's evolution is dictated by the available frequency services (FS) including inertia and FR.

Synchronous thermal generators provide inertial response from the kinetic energy stored in their rotating turbines and FR from governor droop controls. Decarbonisation replaces synchronous thermal generators with renewable energy sources. This means that frequency security is no longer guaranteed by meeting energy demand alone because currently, converter based generation (like wind) is often not able or not incentivised to provide any FS. Thus explicit frequency security constraints must be included in scheduling to assure system stability. This can result in significant wind curtailment due to the minimum generation from thermal plants that are needed to provide FS \cite{Teng2016}, negatively impacting costs and emissions. The increase in operational cost is referred to here as a system's frequency security cost (FSC).

This motivates the investigation of other possible tools at the disposal of system operators to contain the frequency more cost effectively. UFLS results in a step decrease in demand. It is used traditionally as a last resort containment measure after very large outages (non-credible/unsecured events) \cite{SQSS}, and its potential role to support frequency control during normal operation has been overlooked. 

The majority of UFLS literature focuses on how to contain unsecured events with the least load shed \cite{HAESALHELOU2020106054}. This necessitates a fast nadir prediction following the event, from which the optimum load to be shed can be calculated online. If the outage size is known, then the single machine swing equation can be used \cite{Sanaye-Pasand2009}. When outage size is unknown, it can be estimated from initial RoCoF measurements \cite{Terzija2006}. Alternatively, the future frequency evolution can be predicted directly from frequency measurements post disturbance \cite{Rudez2016}. %Continual monitoring of the RoCoF allows improved shedding by considering the impact of FR \cite{Rudez2011}.
Another option is to use artificial neural networks to infer the outage size and subsequent optimum load shed amount \cite{Hooshmand2012}.

All these methods try to minimize the UFLS used post-event, as a containment measure in a system with set FS and commitment decisions. Accurate nadir predictions will implicitly consider system inertia and FR, but offer no framework to compare their cost effectiveness in securing a given outage against UFLS.

In the frequency constrained scheduling literature, %it is common to derive a minimum bound on system inertia based on a linear RoCof constraint derived from the swing equation \cite{Luo2020}. 
the main mathematical challenge lies in containing the frequency nadir, as it necessitates incorporating the dynamic equations dictating post-event frequency evolution into the algebraic optimisation problem \cite{Luo2020}. A highly non-linear nadir constraint considering thermal-plant response only, is derived in \cite{Ahmadi2014} and implemented in the unit-commitment via piece wise linear functions. Reference \cite{Wen2016} augments the formulation of \cite{Ahmadi2014} to consider response from batteries by suggesting a control scheme that effectively reduces the size of the loss. Reference \cite{Paturet2020} derives a nadir constraint incorporating response from grid forming inverters which is used in dynamic simulations to determine the set of secure post-outage plant combinations. Appropriate decision variable are then bounded within the unit commitment to insure commitment lies within this set.

References \cite{Chavez2014,Teng2016,Badesa2019,Teng2017} approximate the FR provision from droop control as a linear ramp to simplify the analytical formulation. %, an approach that is also used here. 
These works propose: a linear constraint on FR from turbines in a system with fixed inertia \cite{Chavez2014}; a linear approximation of the true constraint for systems with variable inertia and single speed FR \cite{Teng2016}; and a reformulation of the nadir constraint as a SOC, resulting in a SOC program that optimises over inertia, a reduces largest loss, fast and slow FR \cite{Badesa2019}. All these papers contain the nadir in order to prevent the triggering of UFLS. 

%The main contribution of this paper is in providing this framework. Breaking the paradigm of UFLS for last resort only, and demonstrating that when considered alongside other FS, UFLS is a valuable resource to contain outages of the single largest power infeed (secured events) in low inertia systems.\textbf{should mention that the framework comes from a FS constraint derived from the swing equation, {signpost what I've done!}, set up the review of SCUC nicely this way.} 

%This comparative framework is necessary to find the optimal FS composition during scheduling. It might lead to lower system cost having to disconnect a small amount of load every 10 years, than it is to schedule additional FR of inertia for that period of time.
    
The only framework that considers UFLS as an option to contain secured events is \cite{Teng2017}, whose nadir constraint bounds the product of the inertia and FR from thermal plants to be larger than a constant. The outage size is assumed constant, and the UFLS amount is discretized into blocks, with one nadir constraint per block. The optimizer can then choose the optimal UFLS level to schedule using binary variables.  A piecewise linear approximation is used to apply this constraint within a mixed-integer linear programme framework.

This paper derives a nadir constraint that allows moderate amounts of UFLS to assist in containing the nadir. The expected cost of UFLS is found probabilistically, as the product of the probability of a largest loss occurring in a given hour and the cost of the load-shed it would initiate. This expected hourly cost facilitates the direct comparison of UFLS to other traditional FS in order to secure frequency most cost effectively. Breaking the paradigm of UFLS as a last resort only and demonstrating that when considered alongside other FS, UFLS is a valuable resource to contain outages of the single largest power infeed (credible/secured events) in low inertia systems. 

The nadir constraint proposed here is significantly more advanced than the linear one of \cite{Teng2017}, as it can consider a dynamically reduced largest loss, and multiple speed FR. Both of which \cite{Badesa2019} has shown to be extremely effective frequency containment measures in low inertia systems, thus vital to consider when accurately evaluating UFLS's value in reducing system FSC. We demonstrate that the proposed nadir constraint can be closely approximated by a single second order cone at each UFLS level. Resulting in a tractable mixed-integer second order cone programme (MISOCP) formulation that fully exploits the capabilities of modern convex optimisation solvers.

%The only framework to do this currently is in \cite{Teng2017}, which augments the linear nadir constraint in \cite{Teng2016} with a step decrease in demand once the frequency security boundary is reached. The UFLS amount is discretized into blocks, with one nadir constraint per block. The optimizer can then choose the constraint enforced using binary variables. Thus balancing a reduced need for FR and inertia in a given timestep against the expected cost of UFLS occurring. However, this MILP framework cannot consider two speeds of frequency response, or the option to reduce the largest loss. Both of which \cite{Badesa2019} has shown to both be extremely effective frequency containment measures. Thus not considering them prevents accurately evaluating UFLS's value in reducing system FSC. This paper's framework has that capability.

The key contributions of this work are:
\begin{enumerate}
    \item A comprehensive framework to co-optimise UFLS along with fast and slow FR, inertia and reduced largest loss,  to secure frequency in the most cost effective manner.
    \item A novel, least conservative convexification of the frequency nadir constraint using SOCs. The computational tractability of this formulation is demonstrated.
    \item The application of the developed model to a representative GB 2030 system.  The sensitivity of UFLS value to: UFLS amount and cost; availability of alternative FS; and wind penetration is investigated.
    %\item A quantification of the value of UFLS in system operation, considering several sensitivities. Probabilistic scheduling of UFLS is shown to be  highly effective for low-inertia power systems.
\end{enumerate}

This paper is organised as follows: Section \ref{SUC} details the stochastic unit commitment (SUC) model used to identify the value of UFLS in annual system operation. Section \ref{FSC} details the formulation of the convex frequency security constraints and section \ref{CV} confirms their veracity with dynamic simulations. Case studies exploring the value of UFLS are presented in Section \ref{CS}, whilst section \ref{Conclusion} gives the conclusion.
\section{Stochastic Unit Commitment Model} \label{SUC}

This paper applies novel frequency security constraints to a multi-stage unit commitment problem to demonstrate the constraints' value in reducing operating costs. The model optimally schedules energy production, reserve and ancillary services in light of uncertain renewable output. 

A single-bus power system model is used despite the fact that that the novel frequency security constraints are applicable to transmission constrained models. Not including transmission constraints, via the dc load flow equations or otherwise, causes system operating costs to be underestimated. However, this is justified because: 1) Frequency is a system wide quantity, so multiple buses do not significantly improve insight into frequency constrained operation, the core contribution of this paper. 2) Operators often solve the unit commitment and energy dispatch problems sequentially, finding nominal plant operating conditions and then correcting for line flows. This paper models the first stage problem.

To capture the critical information about uncertainties, user defined quantiles of the random variable, net demand (demand subtract wind power), are used to construct a scenario tree. The scenarios discretize the continuous range of potential future realisations of the uncertain variable in a representative manner. Reference \cite{Sturt2012} demonstrates that a small number of well selected scenarios, branching at the root node only, can capture most of the benefit of SUC, with high computational tractability. %Figure \ref{tree} shows a typical scenario tree.

%\begin{figure}[t!]
%\centerline{\includegraphics[width=.85\linewidth]{Stochastic Unit_Commitment_Model/Scenariotree.JPG}}
%\caption{Schematic of a typical scenario tree used in SUC \cite{Badesa2019}.}
%\label{tree}
%\end{figure}

The model utilises a rolling planning approach. The operating decisions that minimise the expected cost over the 24hr time horizon are found by using the probability of reaching each scenario as a weighting. It is uneconomical to ensure that for all eventualities demand will be met, so load shedding is permitted in extreme cases. Note, $P^{LS}$ is load shedding from a lack of reserve, and is distinct from $P^{UFLS}$ which is load shedding to assist frequency security, which incurs a cost only when it is activated after a generation outage. 

The SUC objective function is:
\begin{equation}
\label{Cost Function}
   \sum_n^{N} \pi (n) \Bigg( \sum_g^G C_g(n) + \Delta \tau(n) \cdotp c_{LS}  P^{LS}(n) + C^{U}(n)\Bigg)
\end{equation}
The objective function (\ref{Cost Function}) is subject to: typical generator and storage constraints listed in Section III of \cite{Sturt2012}; the power balance constraint (\ref{power balance}) and the frequency security constraints derived in Section \ref{FSC} of this paper. 
\begin{equation}
\label{power balance}
    \sum_g^G P_g(n) + \sum_s^S P_s(n) + P^{LS}(n) + P^w(n)- P^{wc}(n) = P^d(n) 
\end{equation}
The simulation implements the optimal decisions at the here-and-now root node, updating the system states accordingly. At the next time step the actual wind realisation becomes available, from which a new scenario tree is constructed and the process iterated. 

\section{Frequency Security constraints} \label{FSC}
Each node in the scenario tree has a complete set of system decision variables, so the index $n$ is dropped in this section to improve equation clarity. After a generator loss, a system's frequency time-evolution is described by the swing equation \cite{KundurBook}:
\begin{equation}
\label{swng equation}
\frac{2 H}{f_0} \frac{d \Delta f}{d t} = R_1(t) + R_2(t) - PL_{max}
\end{equation}
Equation (\ref{swng equation}) neglects load damping because the level will be significantly reduced in future power-electronic dominated systems \cite{Chavez2014}.  As in \cite{Chavez2014,Teng2016,Teng2017,Badesa2019}, frequency response is modelled as a linear ramp, fully specified by its magnitude and delivery time. Detailed dynamic simulations in Section III of \cite{Badesa2020} show that frequency droop controls can be conservatively and accurately approximated by a ramp. More detailed FR dynamic models would impede the algebraic derivation of convex constraints from the swing equation, as no closed form frequency security conditions could be deduced.
\begin{equation}
\label{Unit ramp}
R_{i}(t) = 
    \begin{cases} 
      \frac{R_{i}}{T_{i}}\cdotp t & t \leq T_{i} \\
      R_{i} & t > T_{i} \\
    \end{cases}
    \, i \in 1,2
\end{equation}
In this paper $T_1 < T_2$. The faster speed can correspond to response from power-electronics devices like batteries, while the slower speed is used to model generator dynamics.

\subsection{RoCoF Constraint}\label{RoCoF Section}
%The maximum RoCoF occurs the instant of generator disconnection, before any FR has been delivered. The maximum RoCoF must be constrained to prevent distributed generation disconnection (which would further exacerbate the imbalance), derived directly from (\ref{swng equation}):
 Relays require several periods to measure the frequency accurately, thus the measured RoCoF is calculated as the average over some time period ($T_s$). The maximum RoCoF occurs in the time period immediately after disconnection, before significant FR has been delivered. It must be limited to prevent the disconnection of distributed generation, via RoCoF-sensitive protection schemes, from exacerbating the demand-generation deficit:
\begin{equation}
\label{RoCoF}
|RoCoF| = \frac{|\Delta f(T_s)|}{T_s} \leq |RoCoF_{max}| 
\end{equation}
Assuming $T_s < T_1$, an expression for $\Delta f(t)$ can be found from (\ref{swng equation}) by integrating between 0 and t:
\begin{equation}
\label{RoCoF.1}
\frac{\Delta f(t)}{t} = \Big( \frac{t}{2T_1} \cdotp R_1 + \frac{t}{2T_2} \cdotp R_2 - PL_{max} \Big) \cdotp \frac{f_0}{2H} 
\end{equation}
Thus, by substituting (\ref{RoCoF.1}) into (\ref{RoCoF}), the linear constraint to limit the maximum RoCoF is found: 
\begin{equation}
\label{RoCoF.2}
 \frac{T_s}{2T_1} \cdotp R_1 + \frac{T_s}{2T_2} \cdotp R_2 + \frac{2 |RoCoF_{max}|}{f_0} \cdotp H \geq PL_{max}
\end{equation}
\subsection{Nadir Constraint with UFLS}
As shown in (\ref{swng equation}) frequency decline occurs when demand is larger than generation. From this it follows that the lowest frequency (nadir) will occur at the instant when the power deficit is made zero. Scheduling at least enough FR to cover $PL_{max}$ insures this will happen before $T_2$:
\begin{equation}
\label{Steady state}
 PL_{max} < R_1 + R_2
\end{equation}%

\begin{figure*}[!b]
%\vspace{-10pt}
\noindent\rule{\textwidth}{0.7pt}
\addtocounter{equation}{+2}
   \begin{equation}
\label{t_hat 2}
\hat{t} = \frac{T_2}{R_2} \cdotp \Bigg( \sqrt{\frac{-4H \Delta f_{trig} R_2 + f_0 R_1 R_2 T_1}{f_0 T_2}    +  (PL_{max}-R_1)^2} + (PL_{max}-R_1)      \Bigg)
\end{equation}
\addtocounter{equation}{+1}
    \begin{equation} 
    \label{non convex SOCP}
    \Bigg(\underbrace{\frac{1}{\sqrt{2}}\Big(\frac{R_2}{2T_2}+\frac{H}{f_0}-\frac{R_1T_1}{4 \Delta f_{trig}} \Big)}_{= \ Z}\Bigg)^2 \geq \Bigg(\underbrace{\frac{1}{\sqrt{2}}\cdotp \Bigg(\frac{R_2}{2T_2} -\frac{H}{f_0} + \frac{R_1 T_1}{4\Delta f_{trig}} \Bigg)}_{= \ X}\Bigg)^2 + \Bigg(\underbrace{\frac{PL_{max} - R_1 }{2 \cdotp \sqrt{\Delta f_{trig}}}}_{= \ Y}\Bigg)^2 - \Bigg(\underbrace{\frac{P_{UFLS}}{2 \cdotp \sqrt{\Delta f_{trig}}} }_{= \ u}\Bigg)^2
    \end{equation}
    \vspace*{-12pt}
\end{figure*}

FR reduces the deficit via power injection from storage or generators. UFLS reduces the deficit via a step demand reduction, initiated when low frequency relays measure the local frequency to have dropped below a predefined threshold \cite{WesternPower}. As such, in real terms scheduling UFLS simply means scheduling a reduced combination of other FS that allows the frequency to drop below the threshold frequency. The magnitude of demand disconnected ($P^{UFLS}$) is equal to the sum of the load through the relays chosen to disconnect upon trigger level breach. This value is discrete by nature. The formulation here allows the system operator to schedule a range of block magnitudes, achievable when relay settings can be updated hourly. The method to choose between varying UFLS amounts is detailed in section \ref{binvars}.

The nadir will occur at the UFLS trigger level if the sum of FR delivered by that time and the load shed amount are larger than the outage:
\addtocounter{equation}{-5}
\begin{equation}
\label{min UFLS}
P^{UFLS} \geq PL_{max} - R_1 - \frac{R_2}{T_2} \cdotp \hat{t}
\end{equation}

Where $\hat{t}$ is the time after outage that $\Delta f(t) = -\Delta f_{trig}$. Equation (\ref{min UFLS}) assumes that $\hat{t}$ occurs within the time interval $T_1:T_2$. A valid assumption because equation (\ref{Steady state}) insures the nadir will always occur before $T_2$. The extremely low inertia required for a system to reach $-\Delta f_{trig} \approx - 0.8$Hz before $T_1 \approx 1$s would violate (\ref{RoCoF.2}) for realistic power system parameters.
%if get push back on this I can just add in the constraint that I explicitly use to prevent this in the model. i.e. for R2=1.8, R1 = 0, UFLS LARGE. Just think it muddies the water for this time round

An expression for $\hat{t}$ can be found from (\ref{swng equation}). First, integrate between 0 and $\hat{t}$ to remove the time derivative: 
\begin{equation}
\label{t_hat 1}
\frac{2 H}{f_0} \cdotp \Delta f(t) =  \frac{R_2}{2T_2} \cdotp t^2 - (PL_{max}-R_1) \cdotp t - \frac{R_1 T_1}{2}
\end{equation}
By rearranging and completing the square, \textit{t} for a given frequency can be found. Thus the time of $\Delta f(t) = - \Delta f_{trig}$ is given by (\ref{t_hat 2}).

By substituting (\ref{t_hat 2}) into (\ref{min UFLS}), the nadir constraint can be formulated:
\addtocounter{equation}{+1}
\begin{equation}
\label{non convex rSOCP}
\hspace{-0.1cm}\bigg(  \underbrace{ \frac{H}{f_0}-\frac{R_1 \cdotp T_1}{4 \Delta f_{trig}} }_{= \ z} \bigg)\hspace{-0.1cm}\underbrace{ \frac{R_2}{T_2} }_{= \ x} \hspace{-0.1cm} \geq \hspace{-0.1cm} \bigg(\underbrace{\frac{PL_{max} - R_1 }{2 \sqrt{\Delta f_{trig}}}}_{= \ y}  \bigg)^2 \hspace{-0.1cm} - \bigg(\underbrace{\frac{P^{UFLS}}{2 \sqrt{\Delta f_{trig}}} }_{= \ u}\bigg)^2
\end{equation}
Where $x,y,z$ are linear expressions of decision variables and $u$ is a constant. Finally, it is useful to rearrange (\ref{non convex rSOCP}) into (\ref{non convex SOCP}) with the following substitutions:
\addtocounter{equation}{+1}
\begin{equation}
    \label{substitutions}
    Z = \frac{1}{\sqrt{2}}(x+z), \ X = \frac{1}{\sqrt{2}}(x-z), \ Y = y
\end{equation}

\subsection{Nadir Constraint Convexification} \label{Convexification section}

\begin{figure}[!h]
\centering
\includegraphics[width = 1\linewidth]{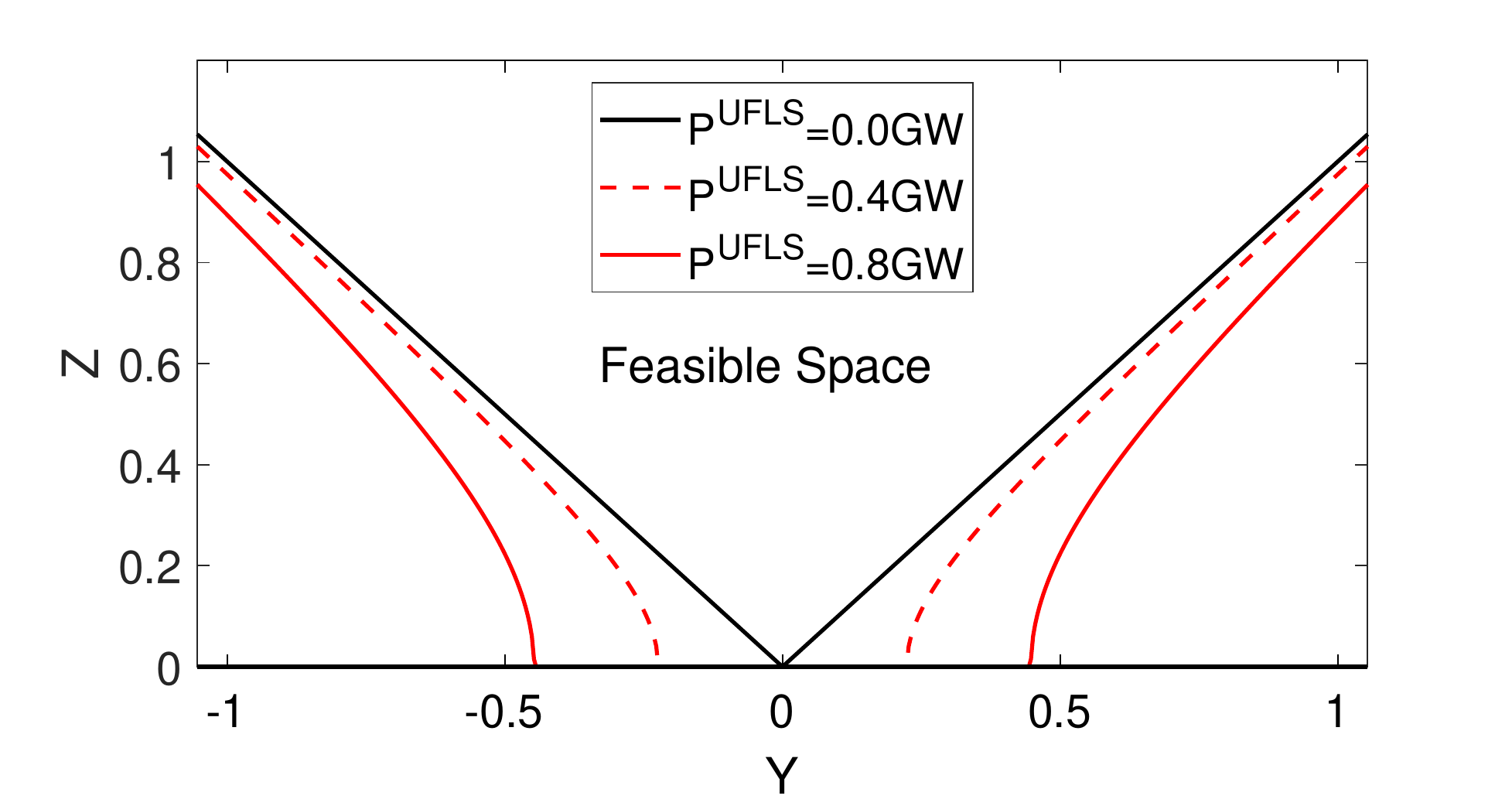}
\caption{Feasible space defined by the nadir constraint (\ref{non convex SOCP}) for various scheduled amounts of UFLS. $Y$ can be considered as an adjusted outage size and $Z$ as an adjusted sum of inertia and FR.}
\label{nonconvex plot}
\end{figure}

The system frequency will not deviate below $-\Delta f_{trig}$ when constraint (\ref{non convex SOCP}) is respected. However, it is non-convex so has limited uses in scheduling and market problems due to low tractability and lack of strong duality properties. The convexification proposed here, stems from the observation that (\ref{non convex SOCP}) is a convex SOC of the classic form $Z^2 \geq X^2 + Y^2$ when there is no UFLS ($u=0$). Fig. \ref{nonconvex plot} shows the constraint over a range of $P^{UFLS}$ values. A 2D plot of the 3D constraint is appropriate because it is symmetrical about any plane that contains the origin and whose normal is perpendicular to the $Z$ axis (including $X=0$).

By considering $Y$ as an adjusted outage size and Z as an adjusted sum of inertia and FR, an intuitive understanding of the constraint's function can be found. As the outage size increases, more FR and inertia is needed to contain the frequency above $(f_0-\Delta f_{trig})$ Hz. UFLS effectively reduces the outage size, thus diminishing FR and inertia requirements and expanding the feasible space.

\begin{figure}[!t]
\centering
\includegraphics[width =1\linewidth]{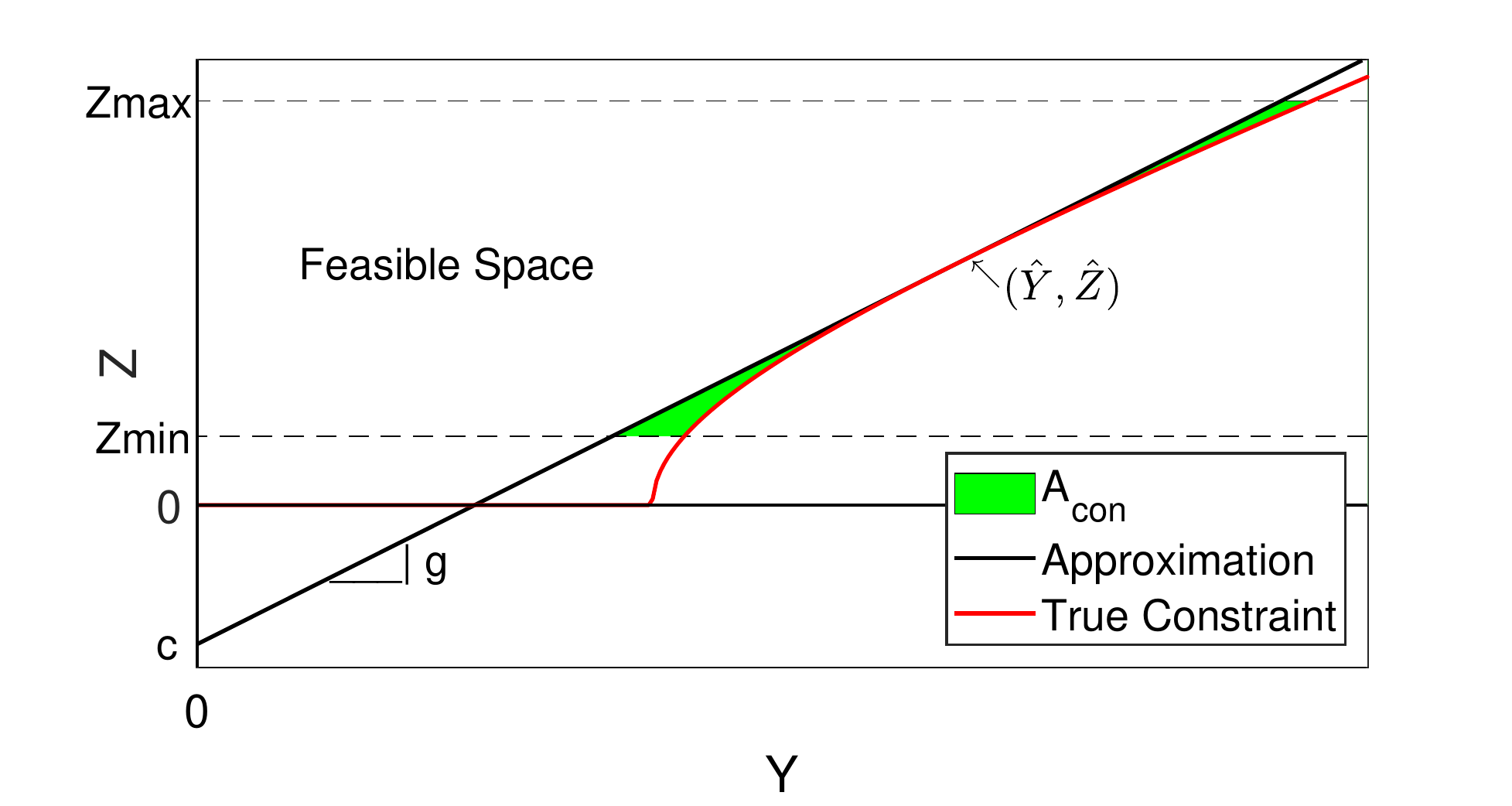}
\caption{Convexification of the nadir constraint (\ref{non convex SOCP}) using a second order cone (\ref{convex least con SOC 2}). For a fixed UFLS amount, the SOC's gradient ($g$) and $Z$ intercept ($c$) are defined by the tangential point ($\hat{Y},\hat{Z}$). The SOC approximation with the minimum $A_{con}$ is the least conservative.}
\label{convexification}
\end{figure}

Although UFLS makes the feasible space nonconvex, the constraint remains similar to a convex SOC, and indeed tends towards the $u=0$ SOC as $Y$ increases, because $Y^2 - u^2 \approx Y^2$ when $Y >> u$. Due to this asymptotic behaviour, any SOC tangential to the nonconvex constraint will never intersect it. Thus the SOC will approximate the true feasible space in a convex and conservative manner, as shown in Fig. \ref{convexification}.  

Solutions that lie on (\ref{non convex SOCP}) are optimal because they secure the frequency with the least FS, and therefor lowest cost. A convex conservative SOC approximation reduces the feasible space, increasing the required FS and cost. The least conservative SOC minimises this feasible space reduction, thus minimising cost increases. For a given $P^{UFLS}$, this is found by minimising the area between the true constraint and the approximation, labelled $A_{con}$ in Fig. \ref{convexification}. We are interested in minimising the reduction in feasible space only, thus $A_{con}$ is further bounded by $Z_{min},Z^{max}$ planes. The Z limits are found by considering the upper and lower bounds of $R_1,R_2,H$. These come from (\ref{RoCoF.2}) and the system specific limits on the provision of these FS from plant and battery levels.

Each point on the constraint's surface has only one pair of SOC gradient ($g$) and zero offset ($c$) that gives the tangential SOC. Thus due to the symmetry about the origin, these are functions of Z. $g$ is the gradient of the constraint (\ref{non convex SOCP}) at the Z value of interest ($\hat{Z}$):

\begin{equation}
\label{Xhat}
\hat{Y}(\hat{Z}) = \sqrt{\hat{Z}^2 + u^2}
\end{equation}
\begin{equation}
\label{g(Zhat)}
g(\hat{Z}) = \frac{\delta Z}{\delta Y}(\hat{Y}(\hat{Z})) =  \frac{\hat{Y}}{\sqrt{\hat{Y}^2 -u^2}}
\end{equation}
Then $c$ is found as the value that a line with gradient $g$ passing through ($\hat{Y},\hat{Z}$) crosses the Z axis:
\begin{equation}
\label{c(Zhat)}
c(\hat{Z}) = \hat{Z} - g(\hat{Z}) \cdotp \hat{Y}
\end{equation}
The $g$ and $c$ that minimise the area $A_{con}(g,c)$ (labelled here as ($g_{opt},c_{opt}$) result in the least conservative SOC approximation because they reduce the feasible space the least. These are found offline by searching over the feasible $\hat{Z}$ range ($Z_{min}~:~Z^{max}$). Thus the new convex, least conservative SOC nadir constraint is given by:
\begin{equation}
\label{convex least con SOC 1}
Z \geq g_{opt} \sqrt{X^2 + Y^2} + c_{opt}
\end{equation}
\begin{equation}
\label{convex least con SOC 2}
\frac{1}{g_{opt}}(Z-c_{opt}) \geq \sqrt{X^2 + Y^2} 
\end{equation}

\subsection{Optimising UFLS amount}\label{binvars}
A constant $P^{UFLS}$ facilitates the convexification, but to optimally balance the UFLS as an option against other FS, an optimizer must be able to choose the amount of UFLS to schedule. Binary variables enable this by successively relaxing any nadir constraint with less UFLS, successively increasing the feasible space and the expected UFLS cost. Assuming:
\begin{equation}
P^{UFLS}_{1}<P^{UFLS}_{2}<...<P^{UFLS}_{N}
\end{equation}
Then the Big M technique can be implemented with (\ref{convex least con SOC 2}) to allow UFLS level choice: 

\begin{subequations}
\label{convex least con SOC 1}
\begin{equation}
Z + M (m_1 + m_2 + ... + m_K ) \geq \sqrt{X^2 + Y^2} 
\end{equation}
\begin{equation}
\frac{1}{\hspace{-0.1cm}g_{opt}(P^{\small{U\hspace{-0.02cm}F\hspace{-0.02cm}L\hspace{-0.02cm}S}}_{1})\hspace{-0.1cm}}(Z-c_{opt}(P^{\small{U\hspace{-0.02cm}F\hspace{-0.02cm}L\hspace{-0.02cm}S}}_{1})) + M (m_2 + ... + m_K) \hspace{-0.1cm} \geq \hspace{-0.1cm} \sqrt{\hspace{-0.1cm}X^2 \hspace{-0.1cm} + \hspace{-0.1cm}Y^2} 
\end{equation}
\hspace{4.25cm}.....
\begin{equation}
\frac{1}{ \hspace{-0.1cm}g_{opt}(P^{UFLS}_{K-1})\hspace{-0.1cm}}(Z-c_{opt}(P^{UFLS}_{K-1})) + M \cdotp m_K \geq \sqrt{X^2 + Y^2} 
\end{equation}
\begin{equation}
\frac{1}{\hspace{-0.1cm}g_{opt}(P^{UFLS}_{K})\hspace{-0.1cm}}(Z-c_{opt}(P^{UFLS}_{K})) \geq \sqrt{X^2 + Y^2} 
\end{equation}
\end{subequations}
\begin{equation}
\label{binary variables sos}
m_1 + m_2 + ... + m_K \leq 1
\end{equation}
The above formulation is a convex SOC because the left hand side remains a linear expression of decision variables. Thus the nadir constraint (\ref{convex least con SOC 1}) can be efficiently solved by any convex solver with MISOCP capabilities, facilitating its tractable application to a wide range of scheduling and market problems. To reduce solver times, it is recommended to keep M as small as possible and to add the binary variables $m_k$ at each node to a special ordered set.

This nadir constraint is an improvement upon the previous most advanced consideration of UFLS  \cite{Teng2017}, which constrains inertia and FR from thermal plants to secure against a fixed loss. We replace the piece-wise linear constraint at each UFLS level with a SOC. Critically, this allows the largest loss size to be a decision variable, and can consider the contributions of two FR speeds alongside UFLS and inertia.

The cost of disconnecting $P^{UFLS}$~GW of load depends on the value of the load that is disconnected ($c^{UFLS}$) and the disconnection length. However, despite being costly when it is utilised, large outages occur rarely so UFLS is seldom triggered, but can substantially reduce the requirements on the other frequency services. To optimally balance these competing effects, the cost of UFLS must be found probabilistically by multiplying the cost of UFLS post-outage by the probability that the outage occurs. This expected cost of UFLS is added to the cost function, and included in the FSC:
\begin{equation}
\label{UFLS expected cost}
C^{U} = t_{rec} \cdotp p \cdotp N_{PL_{max}} \Big( \sum^{K}_{k=1} m_k \cdotp c^{UFLS}_k \cdotp P^{UFLS}_{k} \Big)
\end{equation}
Finally, (\ref{UFLS expected cost}) does not accurately characterise the expected cost of UFLS if any credible losses smaller than $PL_{max}$ would trigger the demand disconnection. This is possible if inertia and FR levels are sufficiently reduced by large UFLS scheduling, that the smaller loss would cause frequency to breach the trigger level. To prevent this the operator must enforce an additional nadir constraint (\ref{non convex SOCP}), with $P^{UFLS}=0$ and $PL_{max}$ equal to the next smallest credible loss. This guarantees that UFLS will only be activated for $PL_{max}$, and thus that (\ref{UFLS expected cost}) reflects the true probabilistic cost of UFLS.

%the true probabilistic cost of . If UFLS is permitted to assist in containing some credible losses smaller than $PL_{max}$ (e.g. a fleet of thermal plants slightly smaller than $PL_{max}$), then an additional nadir constraint (\ref{convex least con SOC 1}) can be enforced for each loss level. With $PL_{max}$ equal to the given loss size. Thus if (\ref{UFLS expected cost}) is augmented with an additional term reflecting the expected UFLS cost from the smaller plants, the true probabilistic cost of UFLS is captured, and the optimal solution can be found. }
\section{Constraint Verification} \label{CV}

\begin{figure}[!t]
\centering
\includegraphics[width=\linewidth]{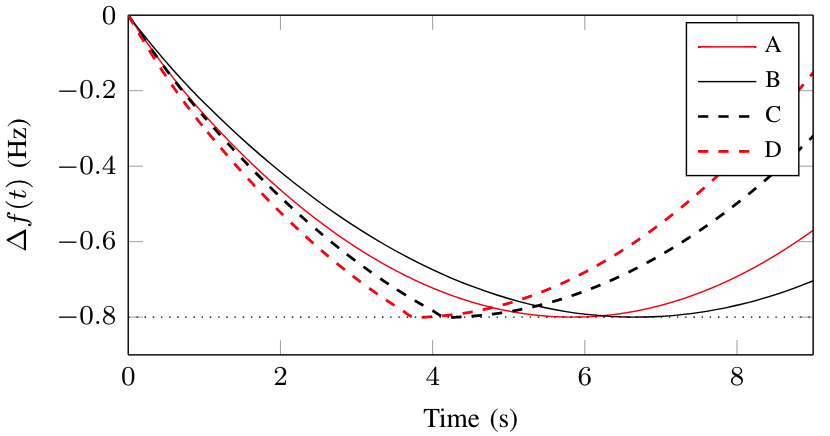}
\caption{Frequency evolution after a loss of $PL_{max}$ for the system conditions in Table~\ref{Dynamic Sim table}.}
\label{Dynamic sims}
\end{figure}

This section uses dynamic simulations from the solution of (\ref{swng equation}), to verify that the nadir constraint (\ref{convex least con SOC 1}) does contain the system frequency, whilst also offering insight into how the different FS interact to do so. Figure \ref{Dynamic sims} plots $\Delta f(t)$ following a $PL_{max}$ outage for the system conditions described in Table \ref{Dynamic Sim table}, all of which lie on (\ref{convex least con SOC 1}). The FR time constants are $T_1=1s, \ T_2=10s$. The nadir will always coincide with the trigger level, because constraint (\ref{convex least con SOC 1}) guarantees that the triggered UFLS is sufficient to close the generation-demand deficit, arresting frequency drop. 

No UFLS is scheduled in cases A and B. Consequently, there is just enough FR and H in both cases to insure that the frequency reaches, but does not cross the UFLS trigger level. However, to secure a loss only 0.2 GW smaller than case B, case A needs 39.1 GWs less  inertia. This is equivalent to the inertia from approximately 20 combined cycle gas turbines (CCGTs). Similarly, comparison between cases C and D shows that fast FR is also an effective FS, with a 0.1 GW difference reducing inertia requirement by 18.6 GWs ($\approx$ 9 CCGTs).

Case C has 20.9 GWs ($\approx$ 10 CCGTs) less inertia than case B. This reduction means that the system frequency breaches the trigger level and UFLS is initiated. The step change in demand reverts the frequency gradient to zero, thus the trigger level corresponds to the nadir. In other words, UFLS has reduced the system's inertia requirements at the expense of an increased UFLS expected cost (\ref{UFLS expected cost}). The constraints developed in this paper offers the optimiser a tool to find the optimal balance between these competing effects during scheduling. 

\begin{table}[!t]
\renewcommand{\arraystretch}{1}
\caption{Dynamic Simulation parameters}
\label{Dynamic Sim table}
\centering
\begin{tabular}{|c|c|c|c|c|c|}
\hline & \hspace{-0.2cm}$P^{\small{U\hspace{-0.02cm}F\hspace{-0.02cm}L\hspace{-0.02cm}}S}$ \footnotesize{(GW)} \hspace{-0.2cm} & \hspace{-0.2cm} $R_1$ \footnotesize{(GW)}\hspace{-0.2cm}& \hspace{-0.2cm} $R_2$ \footnotesize{(GW)}\hspace{-0.8cm}& \hspace{-0.2cm} $H$ \footnotesize{(GWs)} \hspace{-0.2cm}&\hspace{-0.2cm} $PL_{\small{max}}$ \footnotesize{(GW)}\hspace{-0.2cm}\\
\hline
A & 0.0 &	0.2 &	2.4 & 130.7 &	1.6\\
\hline
B & 0.0	& 0.2	& 2.4 & 169.8 & 1.8\\
\hline
C & 0.6	& 0.2 & 2.4 & 146.3 & 1.8\\
\hline
D & 0.6	& 0.3 & 2.4 & 127.7 & 1.8\\
\hline
\end{tabular}
\end{table}
\section{Case Studies} \label{CS}
The SUC model, with cost function (\ref{Cost Function}), was used to identify the savings in annual FSCs from UFLS. Each node is subject to: RoCoF constraint (\ref{RoCoF.2}); minimum FR requirement (\ref{Steady state}); nadir constraint (\ref{convex least con SOC 1},\ref{binary variables sos}); demand-power balance (\ref{power balance}); and the generator and storage constraints listed in section III of \cite{Sturt2012}.

Each simulation corresponds to four months operation of a representative GB 2030 system, one from each season. A scenario tree branching 7 times at the root node only was used to account for net demand forecast uncertainty, as detailed in \cite{Sturt2012}, with quantiles of 0.005,  0.1, 0.3,  0.5, 0.7, 0.9 and 0.995.

A typical time-series of demand was used, with a range of 20:60 GW with daily and seasonal trends. Unless otherwise stated there is an installed wind capacity of 35 GW, and a storage and generation mix detailed in Table \ref{Gen mix}. Frequency security requirements were aligned with GB standard, $f_0=50$~Hz, $|\Delta f_{max}|$ = 0.8~Hz, $RoCoF_{max} $ = 1 Hz/s and $T_s = 0.2$~s (10 cycles at 50~Hz).  The FR time constants are $T_1$ = 1s, $T_2$=10s, while $c_{LS}=$£30,000/MWh and $c^{UFLS}=$£30,000/MWh unless otherwise stated. UFLS was assumed to disconnect load for 1h when utilised, implying that a system re-dispatch would be possible within one hour of the outage. $PL_{max}$ is determined by the power output of nuclear plants, which have an assumed outage rate of 1.8 occurrences/yr. 

The maximum UFLS considered is 0.8 GW, chosen because it is always less than the load shed (5\% of national demand) at the current first UK UFLS trigger level \cite{WesternPower}. The large size difference of at least 0.7 GW, between the nuclear and CCGT plants means that application of an additional nadir constraint, as outlined in Section \ref{binvars}, is unnecessary here. This is because 0.8 GW or less of UFLS never reduces inertia and FR levels to the point where the smaller loss could trigger UFLS.

Simulations were run in an 8 core Intel Xeon 2.40GHz CPU with 64GB of RAM. XPRESS 8.10 was used to solve the optimisations linked to a C++ application via the BCL interface. The MISOCP duality gap was 0.1\%. 
\begin{table}[!t]
\caption{Generation and Storage Characteristics}
\label{Gen mix}
\centering
\renewcommand{\arraystretch}{1}
\begin{tabular}{|l|l|l|l|}
\hline
\textbf{Generation} & Nuclear & CCGT & OCGT \\
\hline
Number of Units & 4 & 120 & 20 \\

Rated Power (GW) & 1.8 & 0.5 & 0.1 \\

Min Stable Generation (GW) & 1.60 & 0.25 & 0.05 \\

No-Load Cost (£'000/h) & 0.0 & 4.5 & 3.0\\

Marginal Cost (£/MWh) & 10 & 47 & 200 \\

Startup Cost (£'000) & NA & 10 & 0\\

Startup Time (h) & NA & 3 & 0 \\

Min up Time (h) & NA & 4 & 0 \\

Inertia Constant (s) & 5 & 4 & 4\\

Max R2 Capacity (GW) & 0.00 & 0.05 & 0.05\\
\hline
\hline
\textbf{Storage} & Pumped & Battery 1 & Battery 2 \\
\hline
Capacity (GWh) & 10 & 1 & 12 \\

Dis/Charge Rate (GW) & 2.6 & 0.3 & 3.0 \\

Max R1 Capacity (GW) & 0.0 & 0.6 & 0.0 \\

Max R2 Capacity (GW) & 0.5 & 0.0 & 0.0 \\

Dis/Charge Efficiency  & 0.75 & 0.95 & 0.95 \\
\hline
\end{tabular}
\end{table}

\subsection{Evaluation of the Conservativeness of Nadir Constraint Approximation}

Equation (\ref{convex least con SOC 1}) conservatively approximates the true non convex feasible space. This produces a highly tractable MISOCP formulation that guarantees system security, at the cost of a reduced feasible space potentially decreasing a solution's optimality. In operational terms, this translates into the over-scheduling of FS. The severity is measured by calculating the minimum nadir that the scheduled FS could secure ($\Delta f_{trig}^{true}$). The closer this value is to $\Delta f_{trig}$, the more optimal the solution, and the better the approximation. The nadir constraint with 0 GW UFLS is not an approximation, so it is not conservative.

\begin{figure}[!t]
\centering
\includegraphics[width=\linewidth]{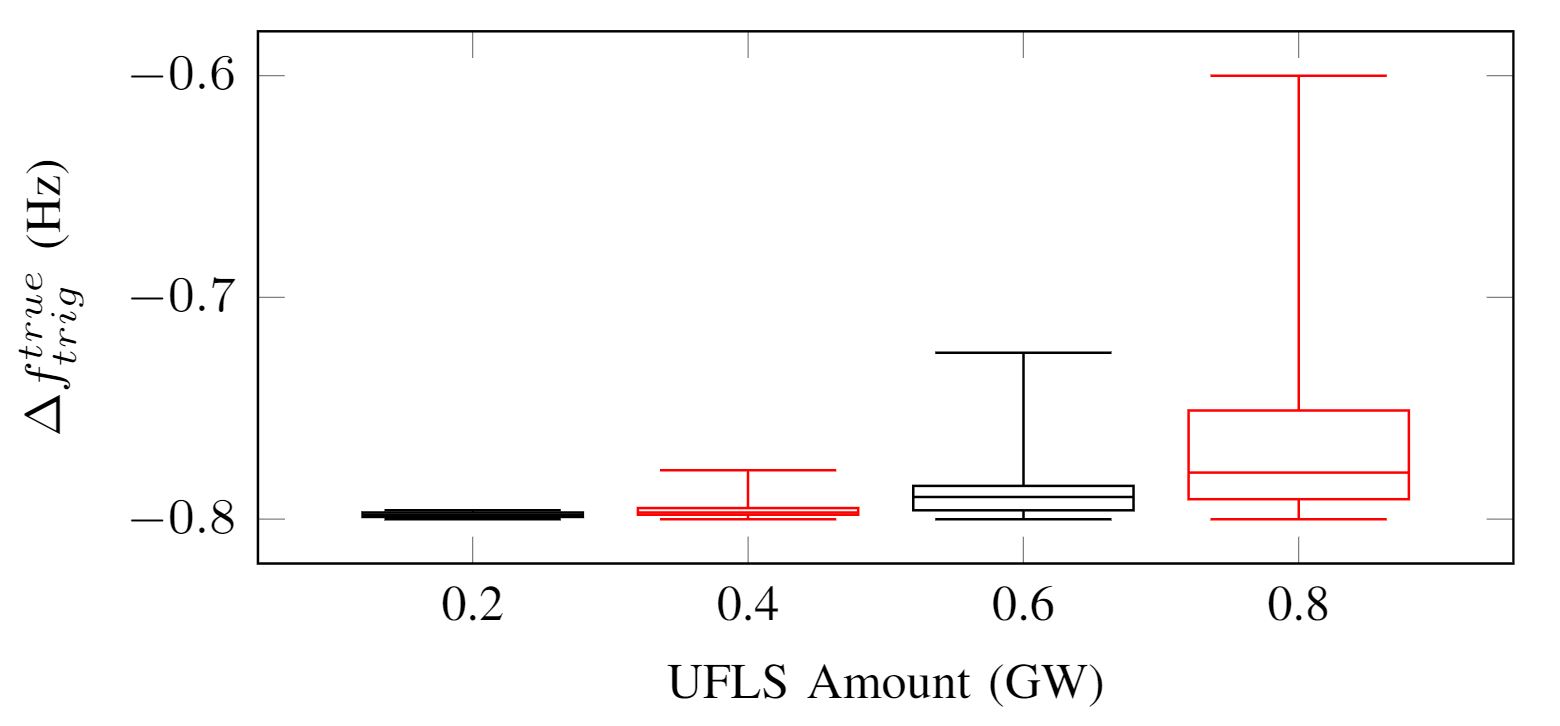}
\caption{The conservative approximation of (\ref{non convex SOCP}) by (\ref{convex least con SOC 1}) results in the over scheduling of FS, which could secure a stricter nadir ($\Delta f_{trig}^{true}$) than that required ($\Delta f_{trig}=0.8$ Hz). Shown here are the annual inter-quartile range, median and max/min of $\Delta f_{trig}$ for systems with two levels of UFLS available, 0 GW or the value shown on the x axis.}
\label{Conservativness}
\end{figure}

Fig.~\ref{Conservativness} shows the distribution of $\Delta f_{trig}^{true}$ for simulations where 2 UFLS levels are available, 0 GW or the value shown on the x axis. For $P^{UFLS}=0.6$ GW, the median $\Delta f_{max}^{true}$ = \mbox{-0.79 Hz}, and the most conservative is -0.73 Hz. Demonstrating that for this level of load shed, the formulation gives solutions that are always less than 10\% conservative, with the vast majority being significantly less so. 

The clear trend of increasing conservativeness with $P^{UFLS}$ in Fig. \ref{Conservativness} can be explained by Fig. \ref{nonconvex plot}.  Larger UFLS levels increase the curvature of the nadir constraint, thus approximating it with a straight line is less accurate. In other words, the minimum area of $A_{con}(g,c)$ becomes larger and the approximation more conservative. However, over the moderate range of available UFLS studied here, the constraint is at the worst 25\% conservative when $P^{UFLS}=0.8$ GW. Even so, when $P^{UFLS}=0.8$ GW the true nadir was below \mbox{-0.7 Hz} over 95\% of the time. Thus most solutions are very close to optimal and insight into the value of UFLS is unhindered. %\textbf{Larger amounts of UFLS are not considered here to avoid activating the nadir constraint for losses smaller than 1 GW.}

\subsection{Value of UFLS in reducing FSC}

\begin{figure}[!t]
\centering
\includegraphics[width=\linewidth]{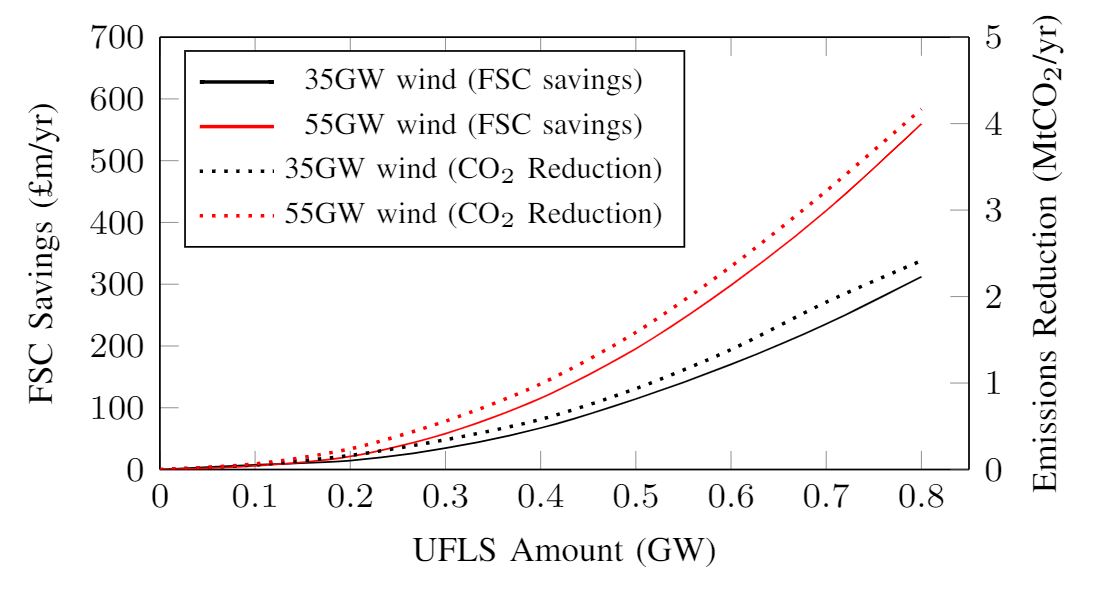}
\caption{The annual reduction in frequency security costs and emissions from optimally scheduling UFLS to support frequency security, in systems with 35 GW and 55 GW of installed wind capacity.}
\label{Co2}
\end{figure}

Fig.~\ref{Co2} shows the reduction in annual FSC when two levels of UFLS are available, 0 GW or the value shown on the x axis. The FSC for a system with 55~GW installed wind capacity and no UFLS is £1072~m/yr, 10.8\% of total system costs. Allowing 0.8 GW of UFLS offers a 52\% reduction of £559m/yr, and 0.4 GW by £115m/yr. These reductions include the increase in expected cost of outages from allowing UFLS, and demonstrate that UFLS can significantly improve system operation in systems with high renewable penetration.

\begin{figure}[!t]
\centering
\includegraphics[width=\linewidth]{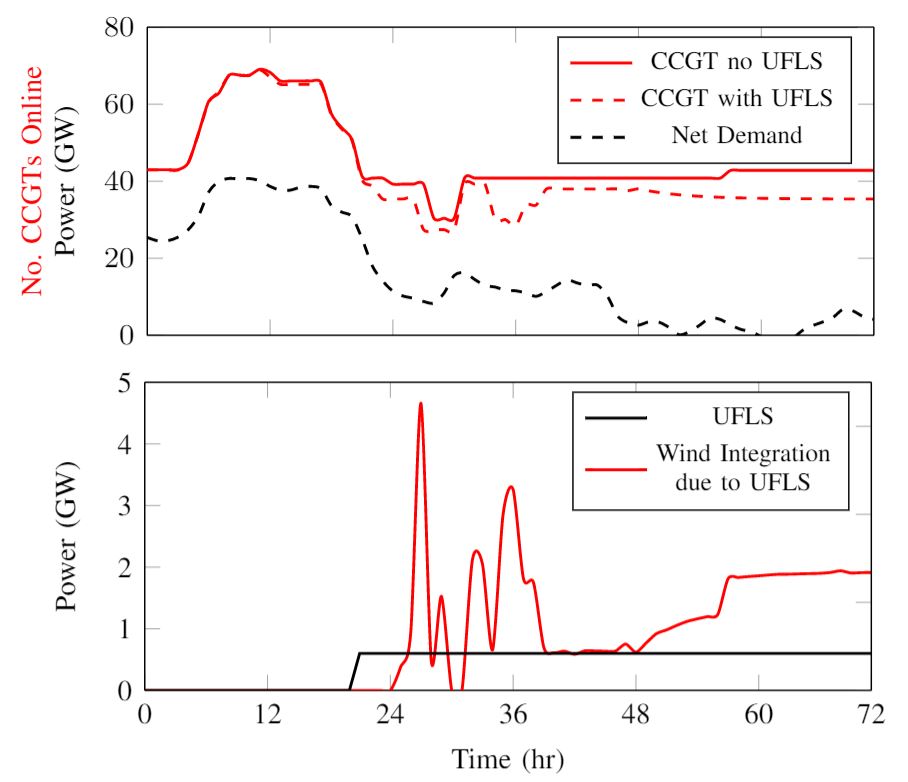}
\caption{Three-day example of the system operation, showcasing the use of UFLS to accomodate more wind. %The top graph shows demand net wind power and CCGT commitment decisions over a 3 day period for identical systems with and without the option to schedule UFLS. The bottom graph shows the optimal UFLS schedule for that 3 day period and the increase in wind power integration it facilitates due to lower thermal plant minimum stable generation.
}
\label{time series}
\end{figure}

The UFLS creates value by reducing the FS required from thermal plants whose minimum generation drives wind curtailment during low net demand periods. An example of this phenomenon is shown in Fig.~\ref{time series}, which compares the optimal commitment of CCGT plants over the same 3-day period for two identical systems. One has the option to schedule 0.6 GW of UFLS and the other does not. CCGTs are shown because nuclear plants have a fixed commitment and the open cycle cycle gas turbine (OCGT) output is negligible over the depicted period. During the high net demand period on the first day, no UFLS is scheduled because frequency security requirements are met as a by-product of energy supply. Accordingly, both systems have similar plants commitment levels.

Overnight, the demand drops and wind output increases, staying high for the following days. Over this period, the scheduled UFLS makes containing the frequency possible whilst needing less inertia and FR from the thermal plants, so fewer CCGTs are run. This lowers the sum of their minimum generation, enabling the absorption of up to 4.7 GWh more wind energy per hour. UFLS's ability to facilitate the absorption of more zero marginal emission wind energy explains the CO2 savings in Fig. \ref{Co2}. Furthermore, when there is higher installed wind generation capacity low net demand periods are more frequent, thus FSC and emission savings increase.

\begin{figure}[!t]
\centering
\includegraphics[width=\linewidth]{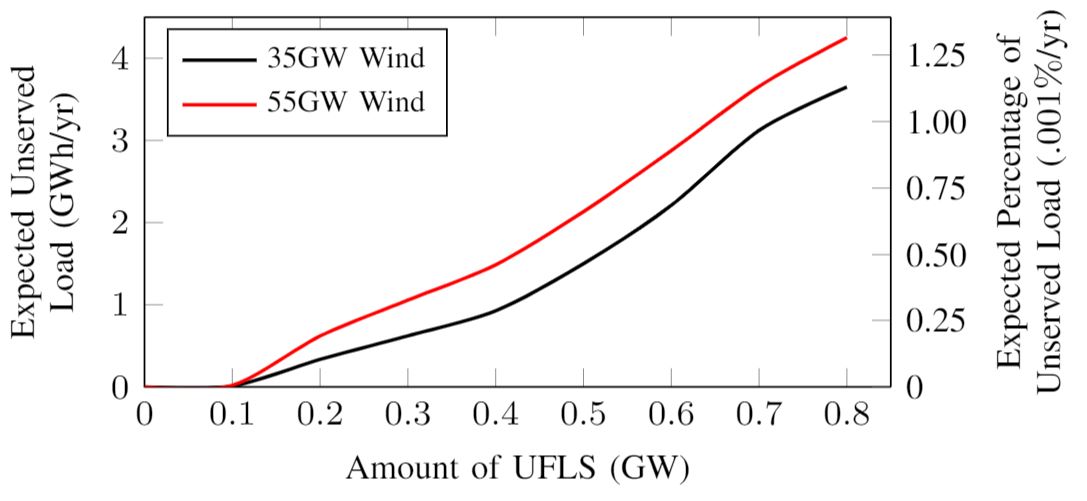}
\caption{The expected amount of annual load shed from optimally scheduling UFLS to assist with frequency security, in systems with 35 GW and 55 GW of installed wind capacity.}
\label{SecuritySupply}
\end{figure}

Load is only disconnected if a nuclear plants outs whilst UFLS is scheduled. The operator risks load shed in order to reduce FR and inertia requirements. The cost savings shown in Fig. \ref{Co2} include the expected increase in load shedding costs. Importantly, the expected amount of load shed to achieve the cost savings is not significant, hence reliability of supply is maintained. For example, 0.8 GW of UFLS in the system with 55 GW of wind decreases FSC by £559m/yr. Fig. \ref{SecuritySupply} shows that the expected load shed to facilitate this is 4.25~GWh/yr, only 0.0013\% of the total load served. All other systems simulated here have lower expected load shed amounts, demonstrating that the proposed framework can assist with frequency security without significantly degrading supply reliability.

%\begin{figure}[!t]
 %   \centering
 % \subfloat[\label{time series 1}]{%
%       \input{Case_Studies/Timeseries.tikz}}
 %     \hfill
 %   \\
 % \subfloat[\label{time series 2}]{%
 %       \input{Case_Studies/timeseries2.tikz}}
%        \hfill
% \caption{(a), (b). Want to collate these into one image}
 % \label{time series} 
%\end{figure}

\subsection{Value of Considering Fast FR and a Reduced Largest Loss}

The proposed framework improves upon the current state of the art nadir constraints. In \cite{Teng2017}, UFLS, inertia and FR from thermal plants are constrained to contain a fixed largest loss. Whereas \cite{Badesa2019} optimally chooses between inertia, fast and slow FR to contain a dynamically controlled loss. Constraint (\ref{convex least con SOC 1}) allows UFLS consideration alongside the other four FS, facilitating the operator to secure frequency in the cost optimal manner.

%It was shown in section \ref{CV} that fast FR ($R_1$) and a dynamically reduced largest loss $PL_{max}$ are highly effective at arresting frequency drop. The failure of \cite{Teng2016} to consider them alongside UFLS inhibits FSC reduction. 

\begin{figure}[!t]
\centering
\includegraphics[width=\linewidth]{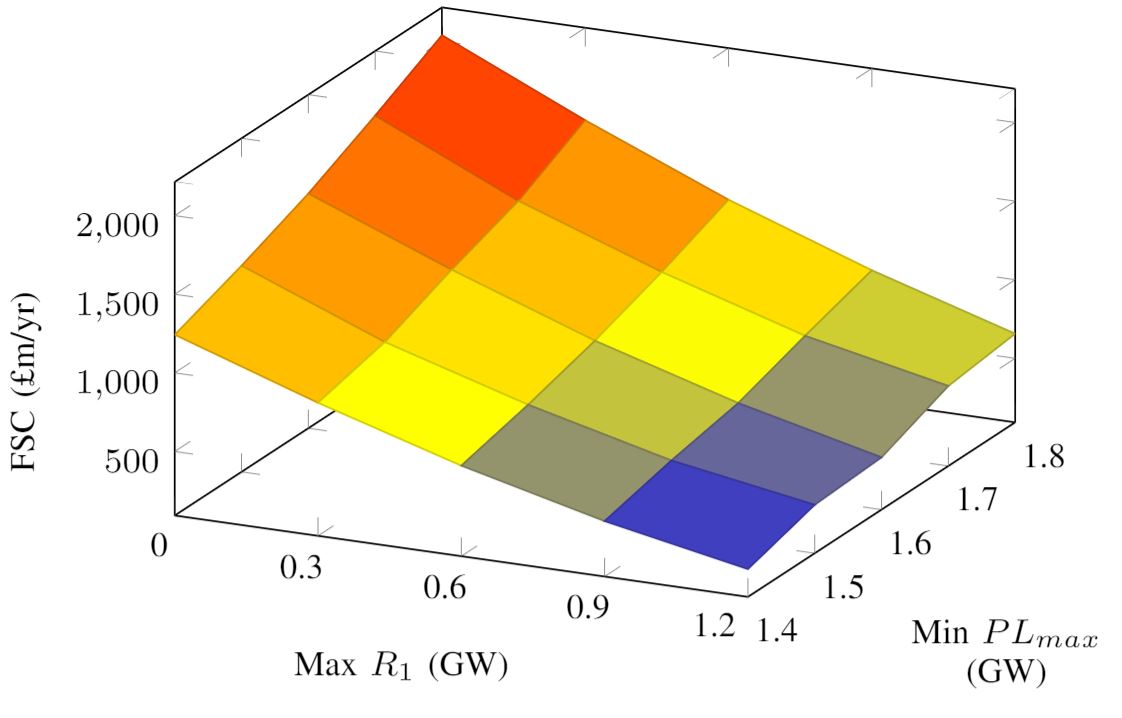}
\caption{Annual FSC for systems with 0.6 GW of UFLS, and varying fast FR and reduced largest loss capabilities.}
\label{FSC Cost Improvements!}
\end{figure}

Figure \ref{FSC Cost Improvements!} shows the total annual FSC for systems with 0.6~GW of UFLS available. The (dis)charge rate of Battery~1 in Table~\ref{Gen mix} was adjusted to simulate varying $R_1$ capacities. The minimum stable generation of nuclear plants was varied to change the lower bound of $PL_{max}$. 

The system with $R_1 \leq 1.2 GW$ and $PL_{max} \geq 1.4 GW$ has a FSC cost of £270m/yr. This is 13.2\% of the £2,040m/yr FSC cost for the system with fixed $PL_{max}=1.8$~GW and $R_1 = 0~GW$. This reduction can be attributed to the high effectiveness of these two services in reducing inertia requirements, as demonstrated in Section \ref{CV}. This lowers plant minimum stable generation, decreasing wind curtailment from 27.3~TWh to 4.3~TWh. Unlike our proposed framework, the constraint of \cite{Teng2017} cannot optimise over these two services, so is inhibited from realising the extensive cost savings they offer.

\begin{figure}[!t]
\centering
\includegraphics[width=\linewidth]{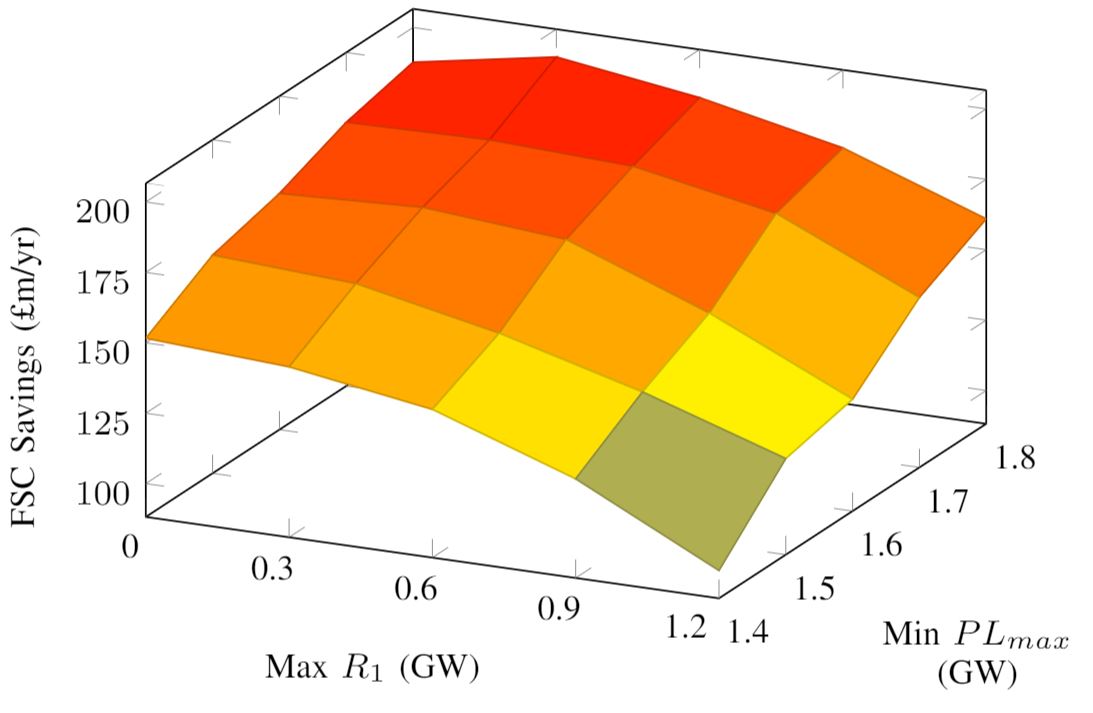}
\caption{Savings in annual frequency security costs enabled by 0.6 GW of UFLS in systems with varying fast FR and reduced largest loss capabilities.}
\label{FSC reduction}
\end{figure}

Fig.~\ref{FSC reduction} shows the reduction in FSC from the ability to schedule 0.6 GW of UFLS, in systems with varying availability of other FS. As the availability of other FS increases, the value of UFLS decreases from £187m/yr in a system with only UFLS, $H$ and $R_2$, to £80m/yr in a system with abundant $R_1$ and $PL_{max}$. This reduction is because the various FS compete to reduce inertia requirements, decreasing each others marginal value. However, UFLS offers significant cost savings even for very flexible systems. Thus, a framework unable to consider it, such as \cite{Badesa2019}, cannot operate a system cost optimally.

Interestingly, Fig. \ref{FSC reduction} shows that when $PL_{max} = 1.8 GW$, the value of UFLS is approximately constant with increasing $R_1$ availability. This is because the nadir requirements at this large loss level cause extreme levels of wind curtailment. There is 28.6~TWh of annual wind curtailment with no UFLS or $R_1$, corresponding to a FSC of £2.23bn/yr, which is 25\% of total system costs. When FSC is this high, additional $R_1$ and UFLS can simultaneously reduce it, without infringing on the other service's marginal value. 

\subsection{Impact of the Expected Cost of UFLS}

\begin{figure}[!t]
\centering
\includegraphics[width=\linewidth]{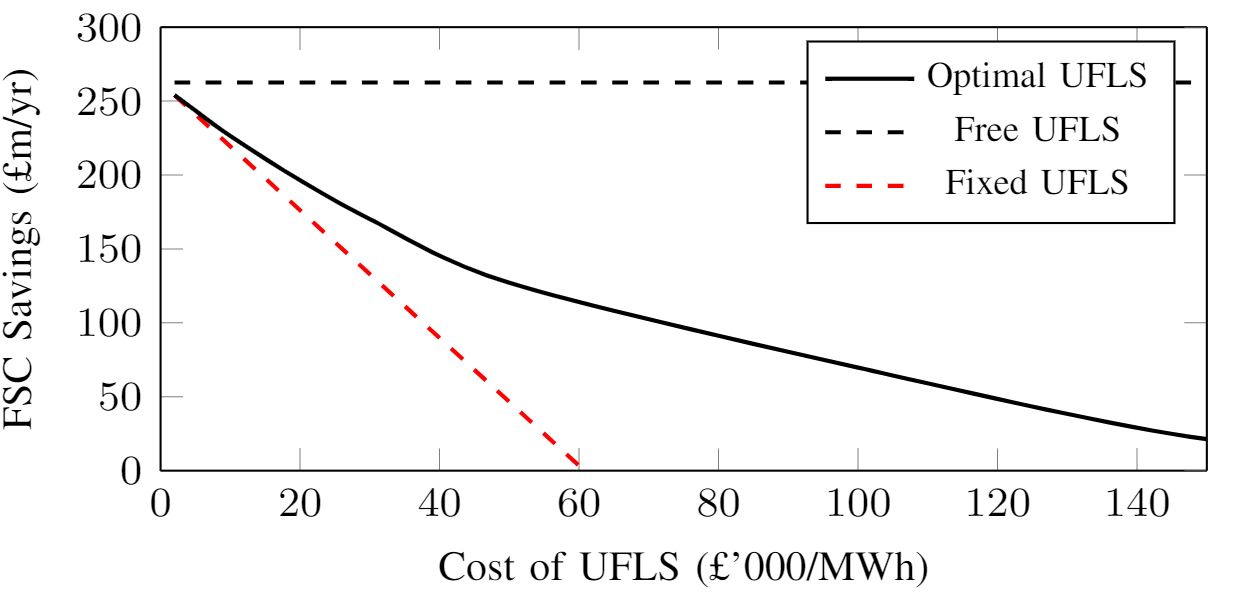}
\caption{Sensitivity of frequency security savings from 0.6 GW of UFLS to the cost of the load shed ($c^{UFLS}$): `Free UFLS' load shed is zero cost; `Fixed UFLS' UFLS is scheduled at all times; `Optimal UFLS' optimiser chooses when to schedule UFLS.}
\label{UFLS cost}
\end{figure}

According to \cite{LondonEconomics2013} the value of load-shed in the UK ranges between 2:45 £'000/MWh depending on the consumer type, importance, season and time of day. Reference \cite{ACER} estimates the range of value of load-shed in mainland Europe to be slightly lower, between 1.3:20 £'000/MWh. In reality it is likely that lower cost loads would be disconnected to support frequency security, so the assumption of $c^{UFLS}$= £30,000/MWh used so far in this paper is a conservative one. Fig.~\ref{UFLS cost} details the sensitivity of UFLS value to $c^{UFLS}$ in 3 separate cases: `Free UFLS' when $c^{UFLS}=0$; `Fixed UFLS' when UFLS is scheduled at all times; `Optimal UFLS' when the optimiser can choose when to schedule UFLS.  

Fig.~\ref{UFLS cost} shows that allowing the scheduling of UFLS creates FSC savings even at extreme UFLS costs of £140,000/MWh, which is more than 3 times the most expensive value of load-shed in the UK. Thus, this paper's framework will provide value when applied to the majority of large power systems. As the expected cost decreases, the penalty for scheduling UFLS decreases, so it is scheduled more frequently and the FSC savings increase. For example, if $c^{UFLS}$ is halved to £15,000/MWh, then Fig. \ref{UFLS cost} shows that the FSC savings from UFLS increase by £41m/yr. 

The expected cost of UFLS ($C^{U}$) is the product of the probability that an outage will occur and the cost of load shedding if it does. As shown in (\ref{UFLS expected cost}), $C^{U}$ is linearly dependant on $c^{UFLS}$, as well as  $t_{rec}, p, N_{Gmax}$. So the sensitivity results in Fig. \ref{UFLS cost} are applicable to adjusting those values also. In other words, if the nuclear plants outed half as often, or the demand was disconnected for half as long, savings of £41m/yr would be observed as those changes are equivalent to reducing the cost of load shedding from £30,000/MWh to £15,000/MWh.

The `Free UFLS' value in Fig. \ref{UFLS cost} corresponds to a system with a permanently reduced FS requirement. As such it represents the maximum FSC reduction from UFLS possible, and is trended towards as $c^{UFLS}$ decreases. On the other hand, as $c^{UFLS}$ increases the `Optimised UFLS' value trends towards 0 because the larger cost of outage makes UFLS a less attractive option to contain the frequency. Above $c^{UFLS}=$£60,000/MWh `Fixed UFLS' shows that always scheduling UFLS begins to increase system costs. This is because the gains from reduced FS requirements, facilitating higher wind absorption, are entirely offset by the severe cost when an outage occurs. The difference between `Fixed UFLS' and `Optimal UFLS' is the value in allowing the optimiser to choose via binary variables (\ref{convex least con SOC 1}), to secure the nadir with UFLS or not.

\subsection{Impact of an Increasing Marginal Load-Shed Cost} \label{margUFLS}

Comparison between case $\alpha$ and $\beta$ in Table~\ref{Variable UFLS Cost} shows the impact of discretising the UFLS into smaller steps. The options to schedule 0.2 GW and 0.4 GW alongside 0 GW and 0.6~GW are introduced. This produces no additional FSC savings, while computation time increased by 68\%. This is because $P^{UFLS}$ has increasing marginal gains, due to its quadratic factor in (\ref{non convex SOCP}), but only linearly increasing costs (\ref{UFLS expected cost}). Accordingly, it is optimal to schedule UFLS levels smaller than 0.6 GW, less than 1\% of the year. This justifies the preceding case studies only considering one UFLS level other than 0 GW. This phenomenon also explains the quadratic increase in FSC savings depicted in Fig. \ref{Co2}. 

However, in reality the marginal $c^{UFLS}$ often increases due to load being shed in the order of value. The formulation can account for this. For case $\gamma$ in table \ref{Variable UFLS Cost}, the first 0.2 GW of scheduled load shed costs only £1,000/MWh, the second 0.2 GW costs £10,000/MWh and the final costs £79,000/MWh. In this case it is optimal to schedule 0.4 GW of UFLS for 48.9\% of the year, a dramatic increase compared to case $\beta$. Explicitly recognising this cost variation facilitates a more optimal system operation, resulting in an annual FSC reduction of £10.6m, despite the fact that the average cost of loadshed is £30,000/MWh when 0.6GW of UFLS is scheduled for both cases $\beta$ and $\gamma$.

\begin{table}[!t]
\renewcommand{\arraystretch}{1}
\caption{Impact of a Variable Marginal UFLS Cost }
\label{Variable UFLS Cost}
\centering
\begin{tabular}{|c|c|c|c|}
\hline
& \begin{tabular}{@{}c@{}} UFLS Steps \\ (GW) \end{tabular} & \begin{tabular}{@{}c@{}} Marginal UFLS Cost \\ (£'000/MWh) \end{tabular} & \begin{tabular}{@{}c@{}} Time at UFLS \\ level (\%) \end{tabular} \\
\hline
$\alpha$ & 0.0, 0.6 & 30 & 49.3, 50.7  \\

$\beta$ & 0.0, 0.2, 0.4, 0.6 & 30, 30, 30 & 50.2, 0.5, 0.3, 49.0 \\

$\gamma$ & 0.0, 0.2, 0.4, 0.6 & 1, 10, 79 & 21.0, 0.2, 48.9, 29.9\\
\hline 
\begin{tabular}{@{}c@{}} \vspace{0.1cm} \\  \end{tabular} & \begin{tabular}{@{}c@{}} FSC Savings \\ (£m/yr) \end{tabular} & \begin{tabular}{@{}c@{}} Run Time \\ (h) \end{tabular} & \multicolumn{1}{|c}{}\\
\cline{1-3}
 $\alpha$ & 170.4 & 5.7 & \multicolumn{1}{|c}{}\\

$\beta$ & 170.4 & 9.6 & \multicolumn{1}{|c}{}\\

$\gamma$ & 180.9 & 7.2 & \multicolumn{1}{|c}{}\\
\cline{1-3}

\end{tabular}
\end{table}

\subsection{Communication Requirements for Scheduling UFLS}
This paper's framework allows the operator to schedule the optimal amount of UFLS from a set of discrete levels. As demonstrated in Section \ref{CV}, UFLS is scheduled by simply scheduling lower amounts of inertia and FR. This causes the post-fault frequency to drop below the UFLS trigger level. The active relays detect this via local frequency measurement, which initiates load disconnection and arrests frequency drop. No communication between the operator and the relays is needed during the fault. However, the net load through relays changes over time, so to maintain a constant UFLS amount (e.g. 0.6 GW) the operator must adjust the set of active relays that will disconnect when the frequency trigger level is breached. This requires an hourly communication between the operator and the relays. When multiple non-zero UFLS levels are available (as in Section \ref{margUFLS}), the set of active relays each hour is chosen so that their net load equals the chosen UFLS level.

\begin{figure}[!t]
\centering
\includegraphics[width=\linewidth]{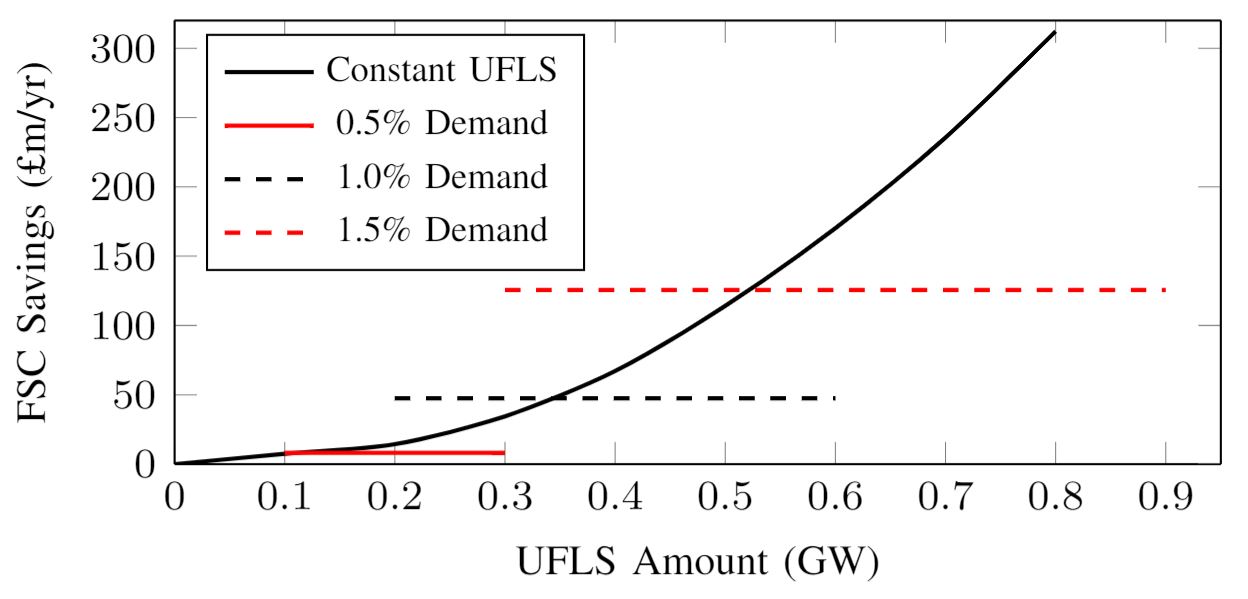}
\caption{FSC savings from two types of UFLS scheme. Constant UFLS, which allows scheduling of a fixed amount of UFLS shown on the x axis, or UFLS that tracks a set percentage of national demand. The range of UFLS levels available within the set percentage schemes is shown by their coverage of the x axis.}
\label{Percentage of demand}
\end{figure}

The presented framework is also compatible with traditional UFLS schemes, that have no communication with relays, and instead disconnect a set percentage of the system demand \cite{WesternPower}. This is done by setting the constant $P^{UFLS}$ in (\ref{non convex SOCP}) equal to the desired fraction of national demand at each timestep, and then calculating $c_{opt}$ and $g_{opt}$ accordingly. Fig.~\ref{Percentage of demand}  compares the annual FSC savings from 0.5\%, 1.0\% or 1.5\% UFLS schemes to those from fixed UFLS amounts. Significant FSC savings of £127m/yr are available from the 1.5\% case, where the available UFLS amounts range between 0.3:0.9~GW. However, this is £187m/yr less than the savings from a constant 0.8~GW of UFLS. The lower average amount of UFLS available contributes to this, but it is compounded by the fact that UFLS is most valuable during periods of low net-demand (as shown in Fig. \ref{time series}), which corresponds to when the set percentage UFLS schemes offer the smallest UFLS amounts.

%The 0.5 GW case offers FSC savings of £115/yr, similar to the 1.5\% case that facilitates an annual FSC saving of £125m/yr. This equivalence, despite the 1.5\% case potentially offering higher amounts of UFLS (between 0.3:0.85~GW), is because the capability of UFLS to reduce required inertia levels is most valuable during low net demand periods. Low demand corresponds to low UFLS availability, thus there is a negative correlation between UFLS abundance and value. This phenomenon is avoided for the constant UFLS case.

The majority of this paper focuses on constant UFLS amounts, despite Fig. \ref{Percentage of demand} demonstrating that the framework is compatible with traditional UFLS systems with no communication infrastructure. There are three main justifications for this:
\begin{enumerate}
    \item For the same or lower maximum UFLS amount, a constant UFLS amount offers significantly more value (up to £187m/yr). Thus there is a large incentive to invest in the requisite communication infrastructure.
    \item UFLS that tracks a set percentage of demand is predicated on the assumption that relays' net demand are proportional to the national demand. This assumption will break down in the future due to the large deployment of distributed generation and active demand (e.g. electric vehicles). Thus frequent communication will be required to maintain a set percentage of demand, which could be better utilised to provide fixed UFLS amounts.
    \item The hourly relay updates required by this methodology are entirely feasible with modern communication equipment. Indeed, the communication requirements for the broad range of adaptive UFLS schemes proposed in the literature \cite{HAESALHELOU2020106054,Sanaye-Pasand2009,Terzija2006,Rudez2016,Hooshmand2012} are much more demanding, often requiring sub-second post fault communications between operator and relays. The proposed framework's local triggering of UFLS makes it robust against communication failures and delays.
\end{enumerate}

\section{Conclusion} \label{Conclusion}
This paper analytically derives a constraint on the frequency nadir from the swing equation. This constraint ensures that there is sufficient fast and slow FR, inertia and UFLS scheduled to contain the largest system loss, which itself can be dynamically reduced to improve operation. We show that when the UFLS amount is discretized, the nonconvex constraint can be accurately and conservatively approximated by a convex second order cone. Case studies were run to explore how scheduled UFLS translates into a reduced requirement on other FS, translating into significantly enhanced integration of wind energy due to the lower response requirements from thermal plants. We demonstrate that drastic cost reductions are achievable by breaking the current paradigm of UFLS as a last resort containment measure. Furthermore the constraint offers insight into the synergies and conflicts between the diverse FS considered. A key take away is that UFLS remains valuable even in highly flexible systems.

There are some areas that could be improved in future. Firstly, this paper focuses purely on how to operate a system at least cost. Availability and utilisation payments for the different FS are not considered. As such, considering UFLS as a service within a market context would allow the investigation of pricing schemes that reflect the appropriate value of different services. Secondly, Fig.~\ref{Conservativness} shows that the nadir constraint is more conservative with increasing UFLS. This is acceptably small for the moderate UFLS amounts investigated here, but could significantly reduce optimality for future systems with abundant UFLS from smart devices. Methods to reduce conservativeness should be investigated, such as: 1) Piece-wise SOC approximation of the true constraint (\ref{non convex SOCP}). 2) Searching within a convex inner approximation of the feasible space around the initial solution from (\ref{convex least con SOC 1}) \cite{Lee2020}. This process can be iterated around each subsequent optimal point until convergence occurs.

Finally, this paper only considers how UFLS can assist with frequency containment after the loss of the single largest plant. However, inertia and FR levels also impact the UFLS needed to contain unsecured events (e.g. the simultaneous disconnection of a nuclear plant and a CCGT). The presented framework could be used to explicitly value their role in containing larger losses, thus facilitating a more cost effective FS procurement.

%, UFLS for unsecured events is not considered separate. However, this framework can consider some unsecured events (e.g. the simultaneous disconnection of a nuclear plant and a CCGT). 

%could enable efficiency gains by blurring this distinction and explicitly considering some unsecured events (e.g. the simultaneous disconnection of a nuclear plant and a CCGT). With the use of a capacity outage probability table, the size and probability of unsecured events can be calculated for use in an additional nadir constraint (\ref{convex least con SOC 1}) at each outage level \textcolor{red}{I wouldn't give so many details on how to plan to expand the work. You should only mention the areas worth investigating, but not say how it should be done (this paragraph was already fine in the 1st submission). I know that reviewer 2 has asked for clarification, but I would do it in a different way, maybe just changing the original sentence in the 1st submission}. This would explicitly value the impact of inertia, FR and UFLS on containing unsecured events, facilitating a more optimal scheduling of these services.

%The formulation will then balance the cost of providing more inertia and FR against the reduced expected UFLS cost when very large outages occur.

%\textcolor{red}{Comments Luis:}
%\begin{itemize}
%    \item \textcolor{red}{If you need space, you can shorten the references by using `et al', removing URLs, just keeping name of journal and year of publication, etc.}
%\end{itemize}

%\appendices
%\section{Appendix Title}

%\section*{Acknowledgement}


\ifCLASSOPTIONcaptionsoff
  \newpage
\fi

\section*{Acknowledgment}
This research has been supported by the UK EPSRC project “Integrated Development of Low-Carbon Energy Systems” (IDLES, Grant EP/R045518/1).

\bibliographystyle{IEEEtran}
\bibliography{Bibliography.bib}
\vspace{-3.2cm}
\begin{IEEEbiography}
    [{\includegraphics[width=1in,height=1.25in,clip,keepaspectratio]{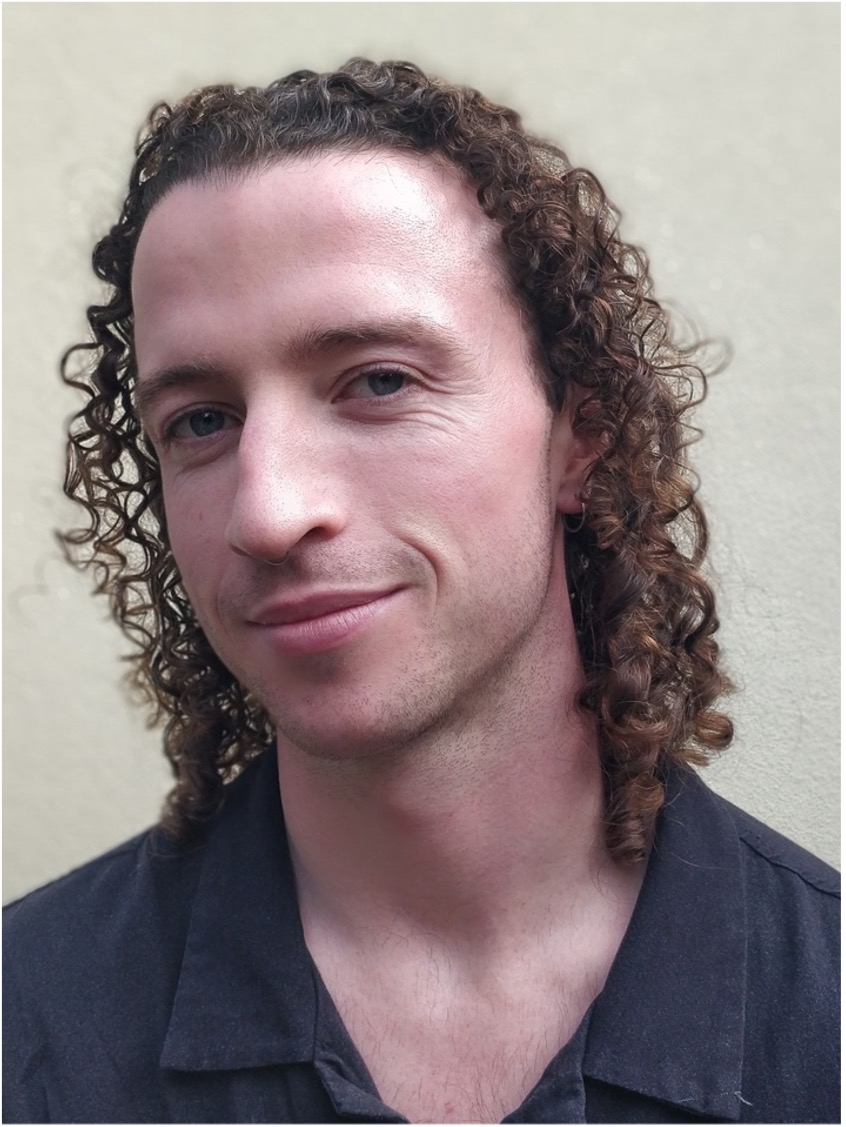}}]{Cormac O'Malley}
(S'18) received the MEng degree in Engineering Science from the University of Oxford, U.K, in 2018. He is currently pursuing a Ph.D. in Electrical Engineering at Imperial College London, U.K. His research interests lie in modelling and optimisation of low carbon power grid operation. 
\end{IEEEbiography}
\vspace{5cm}
\begin{IEEEbiography}
    [{\includegraphics[width=1in,height=1.25in,clip,keepaspectratio]{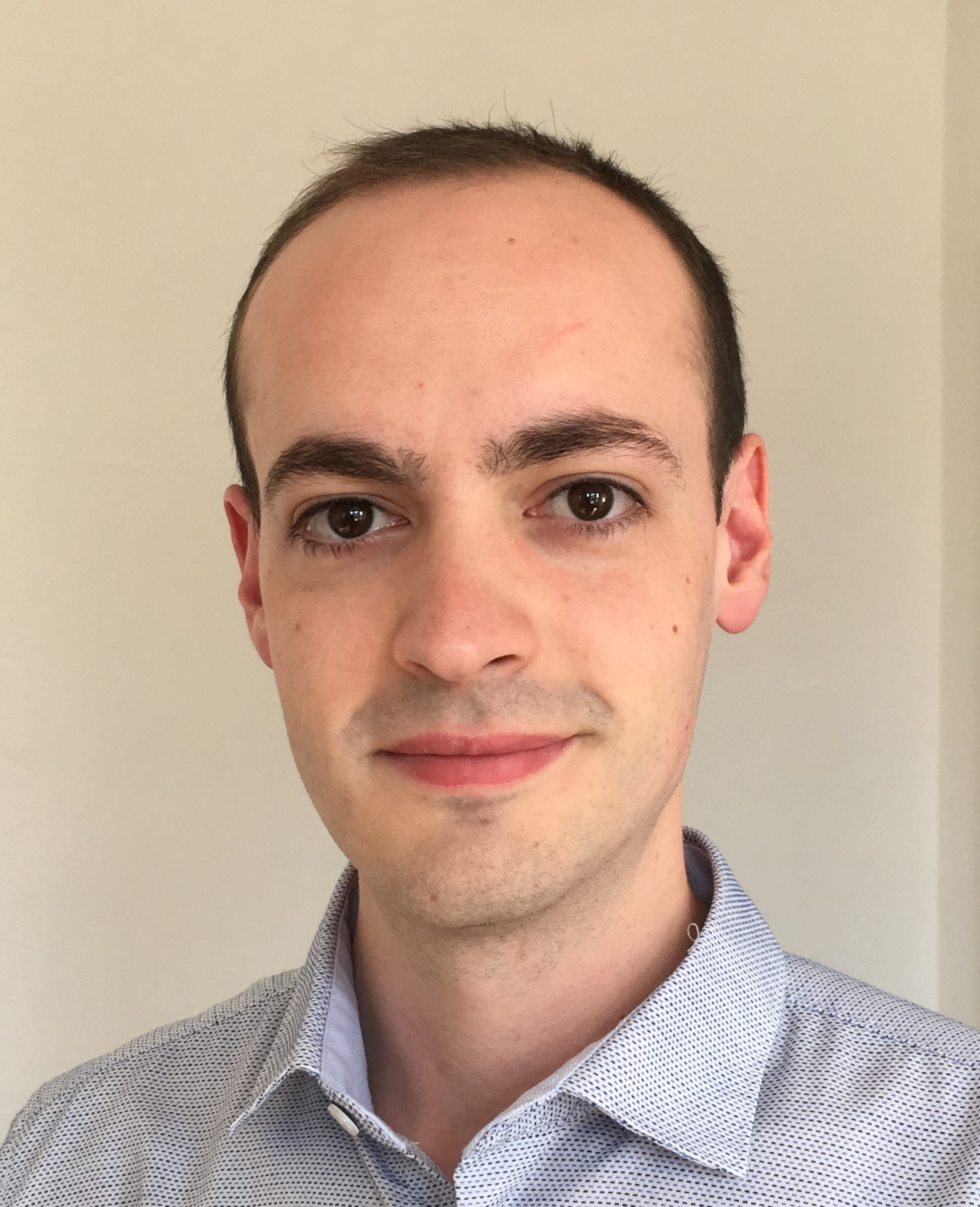}}]{Luis Badesa}
(S'14-M'20) received the Ph.D. degree in Electrical Engineering from Imperial College London, U.K, in 2020. He is currently a Research Associate within the Control \& Power research group at Imperial College London. His research interests lie in modelling and optimisation for low-carbon power grids.
\end{IEEEbiography}
\vspace{-5.5cm}
\begin{IEEEbiography}
    [{\includegraphics[width=1in,height=1.25in,clip,keepaspectratio]{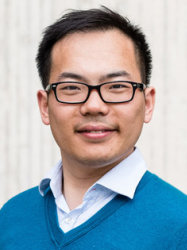}}]{Fei Teng}
(M'15) received the PH.D. degree in Electrical Engineering from Imperial College London, U.K, in 2015. Currently he is a Lecturer in the Department of Electrical and Electronic Engineering, Imperial College London, U.K. His research focuses on scheduling and market design for low-inertia power systems, cyber-resilient energy system operation and control, and objective based data analytics for future energy systems.
\end{IEEEbiography}
\vspace{-4.5cm}
\begin{IEEEbiography}
    [{\includegraphics[width=1in,height=1.25in,clip,keepaspectratio]{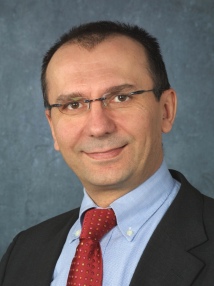}}]{Goran Stbac}
(M'95) is Professor of Electrical Energy Systems at Imperial College London, U.K. His current research is focused on optimisation of operation and investment of low-carbon energy systems, energy infrastructure reliability and future energy markets.
\end{IEEEbiography}

\end{document}